\documentclass[11pt,a4paper]{article}

\usepackage[T1]{fontenc}
\usepackage[utf8]{inputenc}
\usepackage{lmodern}
\usepackage[a4paper,margin=1in]{geometry}
\usepackage{amsmath,amssymb,mathtools}
\usepackage{graphicx}
\usepackage{float}
\usepackage{microtype}
\usepackage{xcolor}
\usepackage{tikz}
\usetikzlibrary{calc}
\definecolor{fpblue}{RGB}{20,76,170}
\definecolor{hored}{RGB}{196,32,32}
\definecolor{ihogreen}{RGB}{0,105,38}
\usepackage[colorlinks=true,linkcolor=blue,citecolor=magenta,urlcolor=blue]{hyperref}

\numberwithin{equation}{section}
\allowdisplaybreaks

\title{Affine Extension of the
Free-Particle--Oscillator--Inverted-Oscillator Triangle}
\date{\small Published in \href{https://doi.org/10.1007/JHEP10(2026)264}{JHEP \textbf{10} (2026) 264}}

\author{
Andrey Alcal\'a\,\href{https://orcid.org/0009-0005-4348-7378}{\textsuperscript{\normalfont\tiny ORCID}}
 and Mikhail S. Plyushchay\,\href{https://orcid.org/0000-0002-1213-1214}{\textsuperscript{\normalfont\tiny ORCID}}\\[3pt]
{\small\itshape Departamento de F\'{\i}sica, Universidad de Santiago de Chile,}\\
{\small\itshape Avenida V\'ictor Jara 3493, Santiago, Chile}\\[3pt]
{\small\slshape E-mails: \textcolor{blue}{andrey.alcala@usach.cl},
\textcolor{blue}{mikhail.plyushchay@usach.cl}}}

\begin{document}

\maketitle

\begin{abstract}
We study the one-dimensional linear-potential  system as an affine
extension of the conformal triangle formed by the free particle,
harmonic oscillator (HO) and inverted harmonic oscillator (IHO). 
Unlike these homogeneous quadratic Hamiltonians in the
\(sl(2,\mathbb R)\) sector, the linear-potential Hamiltonian involves
the Heisenberg ideal of the Schr\"odinger algebra.  Its direct relation
to the free particle is a regular accelerated-frame transformation,
developed here at the levels of the classical action, canonical
transformation, wave-function intertwiner and propagator; its HO and
IHO realizations instead arise through singular displaced-oscillator
limits. The Airy energy eigenstates follow from a cubic-phase transform of
free-particle momentum eigenstates, from condensation of highly excited
harmonic-oscillator levels, and from the limit of a subdominant
parabolic-cylinder scattering branch of the inverted oscillator. 
We also develop a planar extension in homogeneous crossed electric and
magnetic fields, where uniform acceleration generates the electric
interaction, while uniform rotation produces the Landau coupling and
the centrifugal inverted-oscillator term; the guiding-center dynamics
yields the Hall drift.  
\end{abstract}

\section{Introduction}
\label{sec:introduction}

The free particle (FP) and harmonic oscillator (HO) provide one of the simplest
examples of distinct systems governed by the same nonrelativistic conformal
structure.  The maximal kinematical symmetry of the free Schr\"odinger equation
contains an $SL(2,\mathbb R)$ group acting projectively on time, and the
oscillator has an isomorphic symmetry realized by the Cayley--Niederer
transformation \cite{Niederer1972,Hagen1972,Niederer1973}.  In canonical
language, the Hamiltonian, dilation and expansion generators span
$sl(2,\mathbb R)\cong sp(2,\mathbb R)$; their parabolic, elliptic and hyperbolic
representatives are the FP, the HO and the inverted harmonic oscillator (IHO).

The stationary counterpart of the Cayley--Niederer map is the conformal bridge
transformation (CBT), developed for a variety of conformal systems
\cite{InzunzaPlyushchayWipf2020,InzunzaPlyushchay2021Review}.  In two recent
works we combined the time-dependent and stationary viewpoints: first through
projective time, Cayley transformations, the Schwarzian derivative and
metaplectic lifts in the FP--HO correspondence
\cite{AlcalaPlyushchay2026Projective}, and then in the FP--HO--IHO triangle and
its $\mathfrak{osp}(1|2)$ structure \cite{AlcalaPlyushchay2026Triangle}.  Since
the three self-adjoint Hamiltonians have inequivalent spectra and spectral
realizations, these bridges relate distinct realizations of a common conformal
module rather than unitarily equivalent Hamiltonians.

The present paper determines how the linear-potential (LP) system fits into
this picture.  Although a constant force can be removed by a uniformly
accelerated frame, $H_{\rm LP}=H_0+Fq$ is not another homogeneous quadratic
element of $sl(2,\mathbb R)$: its linear term belongs to the Heisenberg ideal of
the Schr\"odinger algebra.  This motivates an \emph{affine Schr\"odinger
bridge}: the direct FP--LP map is a regular Heisenberg--Weyl displacement with
its Bargmann phase \cite{Bargmann,LeviLeb}, while the projective
$SL(2,\mathbb R)$ action is inherited through dynamical integrals.

We construct the regular FP--LP bridge at the Hamiltonian and action
levels, as a canonical transformation and wave-function intertwiner,
and at the propagator level.  We then analyze the associated spectral
transformations and contrast the fixed-\(F\) HO--LP and IHO--LP limits
with this regular accelerated-frame map.
The LP Airy eigenstates arise from a cubic-phase transform of FP
momentum eigenstates, from spectral condensation of highly excited
displaced-HO states, and from a subdominant IHO parabolic-cylinder
branch without level condensation. In contrast to the
nonspreading accelerating Airy packets of Ref.~\cite{BerryBalazs1979}, our
emphasis is on stationary spectral states and their bridge transformations.

A planar extension in homogeneous crossed electric and magnetic fields gives a
further interpretation.  A uniformly accelerated and rotating frame produces
the electric interaction, the Coriolis/Landau coupling and the centrifugal IHO
term; a compensating harmonic term isolates the Landau sector.  In
guiding-center variables the electric field acts linearly on a noncommutative
Heisenberg plane, generates the Hall drift, and supplements the Landau CBT by a
displacement, a uniform drift and a scalar phase.

Section~\ref{sec:LP-algebraic-position} locates LP in the Schr\"odinger algebra
and relates it to the FP--HO--IHO triangle.  Sections~\ref{sec:affine-bridge} and
\ref{sec:LP-propagator-limits} develop the regular FP--LP bridge and the singular
oscillator limits.  Section~\ref{sec:Airy-transform} treats the Airy transform
and spectral limits, Sec.~\ref{sec:crossed-field-extension} the crossed-field
extension, and Sec.~\ref{sec:discussion-outlook} the discussion and outlook.
The IHO--LP spectral limit is completed in the Appendix.
\section{Algebraic position of the linear potential}
\label{sec:LP-algebraic-position}

The free particle, harmonic oscillator and inverted harmonic oscillator are
organized by the standard quadratic conformal generators
\begin{equation}
\begin{gathered}
\{q,p\}=1,
\qquad
H_0=\frac{p^2}{2m},
\qquad
D=\frac12 qp,
\qquad
K=\frac{m}{2}q^2,
\\
\{D,H_0\}=H_0,
\qquad
\{D,K\}=-K,
\qquad
\{H_0,K\}=-2D .
\end{gathered}
\label{eq:sl2-PB}
\end{equation}
Their parabolic, elliptic and hyperbolic representatives are, respectively,
\begin{equation}
H_{\rm FP}=H_0,
\qquad
H_{\rm HO}=H_0+\omega^2K,
\qquad
H_{\rm IHO}=H_0-\Omega^2K .
\label{eq:FP-HO-IHO-Hamiltonians}
\end{equation}
All three Hamiltonians therefore belong to the same homogeneous quadratic
$sl(2,\mathbb R)$ sector.

The linear-potential Hamiltonian,
\begin{equation}
H_{\rm LP}=\frac{p^2}{2m}+Fq,
\label{eq:H-LP-basic}
\end{equation}
generates $\dot q=p/m$ and $\dot p=-F$.  Its basic dynamical integrals are
\begin{equation}
P_F=p+Ft,
\qquad
X_F=q-\frac{t}{m}p-\frac{F}{2m}t^2,
\qquad
\{X_F,P_F\}=1 .
\label{eq:PF-XF}
\end{equation}
Together with the central mass, $P_F$ and $G_F=mX_F$ generate translations and
Galilean boosts, with $\{G_F,P_F\}=m$.  In terms of the conserved pair,
\begin{equation}
H_{\rm LP}=\frac{P_F^2}{2m}+FX_F .
\label{eq:H-LP-XP}
\end{equation}
Thus $H_{\rm LP}$ is a quadratic free-particle element supplemented by a linear
Heisenberg--Galilei generator, rather than another quadratic representative of
$sl(2,\mathbb R)$.

The quadratic combinations
\begin{equation}
H_0^{(F)}=\frac{P_F^2}{2m},
\qquad
D_F=\frac12X_FP_F,
\qquad
K_F=\frac{m}{2}X_F^2
\label{eq:LP-quadratic-integrals}
\end{equation}
are dynamical integrals of the LP system \eqref{eq:H-LP-basic} and
obey the same \(sl(2,\mathbb R)\) Poisson-bracket relations as
\eqref{eq:sl2-PB}.
The natural algebraic setting is consequently the non-semisimple
Schr\"odinger algebra, equivalently the one-dimensional Jacobi algebra
\cite{BernSch},
\begin{equation}
\mathfrak{sch}(1)=sl(2,\mathbb R)\ltimes\mathfrak h_1,
\qquad
H_{\rm LP}\in\mathfrak{sch}(1),
\qquad
H_{\rm LP}\notin sl(2,\mathbb R),
\label{eq:HLP-in-Sch-not-sl2}
\end{equation}
where $\mathfrak h_1$ is generated by $X_F$, $P_F$ and the central unit.  The
Heisenberg algebra is also a dynamical symmetry of FP, HO and IHO, but their
Hamiltonians themselves remain entirely within the quadratic sector.  This is the
sense in which LP is an affine--Heisenberg companion, rather than a fourth
quadratic vertex, of the FP--HO--IHO triangle.

The direct FP--LP map keeps \(\tau=t\) and transports the free conformal
generators into \eqref{eq:LP-quadratic-integrals}; their finite action on time
remains projective,
\begin{equation}
t\longmapsto t'=\frac{at+b}{ct+d},
\qquad ad-bc=1,
\label{eq:LP-inherited-projective-time}
\end{equation}
with global completion on $\mathbb{RP}^1$.  The distinction from the
Cayley--Niederer maps lies in the affine--Heisenberg intertwiner, not in the
absence of projective symmetry.

\subsection{Affine extension of the quadratic triangle}
\label{subsec:affine-extension-family}

The relation to the oscillator sectors is summarized by
\begin{equation}
H_{\kappa,F}=\frac{p^2}{2m}+\frac{m\kappa}{2}q^2+Fq
=\frac{p^2}{2m}+\frac{m\kappa}{2}Q^2-\frac{F^2}{2m\kappa},
\qquad
Q=q+\frac{F}{m\kappa},
\quad \kappa\ne0 .
\label{eq:H-kappa-F}
\end{equation}
For \(F=0\), the choices \(\kappa=0,\omega^2,-\Omega^2\) reproduce FP, HO
and IHO, respectively.  For $\kappa\ne0$, the linear term only displaces the
oscillator center and adds a constant.
The limit \(\kappa\to0\) is regular in the original family
\(H_{\kappa,F}\), where it yields \(H_{\rm LP}\), but it is singular in
the displaced-oscillator representation: both the displacement and the
additive constant diverge.  Thus LP is attached regularly to the
free-particle vertex through the affine accelerated-frame bridge,
whereas its HO and IHO realizations require controlled singular
displaced-oscillator limits.

Noncommuting contractions along two paths in the
\((\kappa,F)\)-plane to the common FP endpoint
\((\kappa,F)=(0,0)\) were studied in
Refs.~\cite{DaboulWolf2006,DaboulPogosyanWolf2007}.
Reference~\cite{DaboulWolf2006} primarily treated the
symmetry-algebra contractions for attractive and repulsive
oscillators, whereas Ref.~\cite{DaboulPogosyanWolf2007} analyzed the
attractive-oscillator wave-function limits along the two paths,
including the HO--LP--FP path and its deformation--contraction
hysteresis. These works concern ordered contractions toward FP.  Here, by
contrast, \(F\) is kept fixed in the singular HO--LP and IHO--LP
limits, the latter involving a parabolic-cylinder-to-Airy
degeneration.  We incorporate these limits at the Hamiltonian,
action, dynamical-integral and propagator levels into the same affine
framework as the regular time-dependent FP--LP bridge, for which the
canonical transformation and wave-function intertwiner are
constructed explicitly. 

The same family admits a concise Newton--Hooke interpretation.
For the potential per unit mass
\(\mathcal U_{\kappa,F}(q)=\kappa q^2/2+(F/m)q\), one has
\begin{equation}
-\mathcal U_{\kappa,F}'=-\kappa q-\frac{F}{m},
\qquad
\mathcal U_{\kappa,F}''=\kappa,
\qquad
q_{\rm c}=-\frac{F}{m\kappa} .
\label{eq:NH-acceleration-curvature}
\end{equation}
Thus $\kappa$ fixes the Newton--Hooke tidal curvature, whereas $F/m$ is a
homogeneous acceleration that shifts the distinguished center without changing
that curvature.
Besides the ordinary flat limit \(\kappa\to0\) at fixed \(q_{\rm c}\),
which gives FP, there is the correlated limit
\begin{equation}
\kappa\to0,
\qquad
|q_{\rm c}|\to\infty,
\qquad
-m\kappa q_{\rm c}=F\quad\text{fixed},
\label{eq:NH-correlated-flat-limit}
\end{equation}
which leaves a finite homogeneous field and produces LP.  In this sense LP is the
uniformly accelerated remnant of displaced Newton--Hooke dynamics in a singular
zero-curvature limit.

At the quantum level all three non-free systems possess natural length scales,
\begin{equation}
\ell_F=\left(\frac{\hbar^2}{2m|F|}\right)^{1/3},
\qquad
\ell_\omega=\sqrt{\frac{\hbar}{m\omega}},
\qquad
\ell_\Omega=\sqrt{\frac{\hbar}{m\Omega}} .
\label{eq:Airy-length}
\end{equation}
The distinction between LP and HO/IHO is therefore algebraic rather than simply
the presence or absence of a quantum scale.

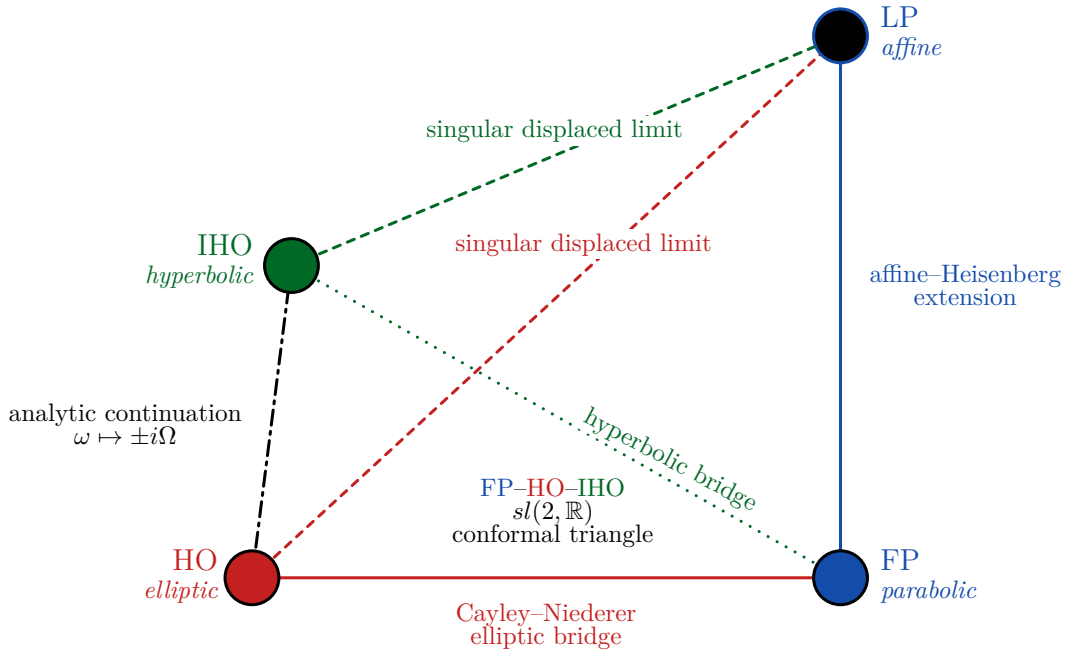
\begin{figure}[H]
\centering
\resizebox{0.88\textwidth}{!}{%
\begin{tikzpicture}[
  x=1cm,y=1cm,
  line cap=round,
  line join=round,
  every node/.style={font=\small},
  vertex/.style={circle,draw=black,very thick,minimum size=7.5mm,inner sep=0pt},
  edgelabel/.style={fill=white,inner sep=1.4pt,align=center,font=\small}
]

\coordinate (HO)  at (0,0);
\coordinate (FP)  at (8.2,0);
\coordinate (IHO) at (0.55,4.35);
\coordinate (LP)  at (8.2,7.55);

\draw[hored,very thick] (HO) -- (FP);
\draw[
  ihogreen,
  very thick,
  dash pattern=on 0pt off 4pt,
  line cap=round
] (IHO) -- (FP);
\draw[black,very thick,dash pattern=on 6pt off 2.5pt on 1.2pt off 2.5pt] (HO) -- (IHO);
\draw[fpblue,very thick] (FP) -- (LP);

\draw[hored,very thick,dashed] (HO) -- (LP);
\draw[ihogreen,very thick,dashed] (IHO) -- (LP);

\node[vertex,fill=hored]     at (HO)  {};
\node[vertex,fill=fpblue]    at (FP)  {};
\node[vertex,fill=ihogreen]  at (IHO) {};
\node[vertex,fill=black,draw=fpblue] at (LP)  {};

\node[anchor=east,align=right,text=hored,font=\large] at ($(HO)+(-0.34,0.03)$)
  {HO\\[-1mm]{\small\itshape elliptic}};
\node[anchor=west,align=left,text=fpblue,font=\large] at ($(FP)+(0.42,0.02)$)
  {FP\\[-1mm]{\small\itshape parabolic}};
\node[anchor=east,align=right,text=ihogreen,font=\large] at ($(IHO)+(-0.40,0.06)$)
  {IHO\\[-1mm]{\small\itshape hyperbolic}};
\node[anchor=west,align=left,text=fpblue,font=\large] at ($(LP)+(0.42,0.04)$)
  {LP\\[-1mm]{\small\itshape affine}};

\node[edgelabel,text=hored,below=3mm] at ($(HO)!0.50!(FP)$)
  {Cayley--Niederer\\[-1mm] elliptic bridge};

\node[edgelabel,text=ihogreen,rotate=-28] at ($(IHO)!0.69!(FP)+(0,0.28)$)
  {hyperbolic bridge};

\node[edgelabel,text=black,anchor=east] at ($(HO)!0.49!(IHO)+(-0.35,0)$)
  {analytic continuation\\[-1mm] $\omega\mapsto\pm i\Omega$};

\node[edgelabel,text=fpblue,anchor=west] at ($(FP)!0.54!(LP)+(0.35,0)$)
  {affine--Heisenberg\\[-1mm] extension};

\node[edgelabel,text=hored] at ($(HO)!0.57!(LP)+(-0.05,0.34)$)
  {singular displaced limit};

\node[edgelabel,text=ihogreen] at ($(IHO)!0.48!(LP)+(0,0.34)$)
  {singular displaced limit};

\coordinate (C) at ($(HO)!0.333!(IHO)!0.50!(FP)$);
\node[align=center,font=\small] at ($(C)+(0,0.18)$)
  {{\textcolor{fpblue}{FP}--\textcolor{hored}{HO}--\textcolor{ihogreen}{IHO}}\\[-1mm]
   {$sl(2,\mathbb R)$}\\[-1mm]
   {conformal triangle}};

\end{tikzpicture}%
}
\caption{Prism-like representation of the LP affine extension of the
FP--HO--IHO conformal triangle.  The lower face contains the parabolic (FP),
elliptic (HO) and hyperbolic (IHO) quadratic representatives.  The vertical
FP--LP edge is the regular affine--Heisenberg bridge, whereas the dashed
LP--HO and LP--IHO edges are singular displaced-oscillator limits with centers
sent to infinity.  The dotted FP--IHO rear edge and dash-dotted HO--IHO edge
denote the hyperbolic bridge and analytic continuation, respectively.}
\label{fig:LP-affine-extension-triangle}
\end{figure}

The regular and singular relations among the four systems are
summarized in Fig.~\ref{fig:LP-affine-extension-triangle}.

\section{The affine accelerated-frame bridge}
\label{sec:affine-bridge}

The direct FP--LP bridge is the time-dependent spatial translation
\begin{equation}
Q=q+\frac{F}{2m}t^2,
\qquad
\tau=t .
\label{eq:FP-LP-coordinate-map}
\end{equation}
Unlike the FP--HO and FP--IHO Cayley--Niederer maps \cite{Niederer1973,AlcalaPlyushchay2026Projective,AlcalaPlyushchay2026Triangle}, it involves no time
reparametrization and is simply the passage to a uniformly accelerated frame.
For
$L_0=m\dot Q^{\,2}/2$ and $L_{\rm LP}=m\dot q^{\,2}/2-Fq$, one finds
\begin{equation}
L_0=L_{\rm LP}+\frac{d\Phi}{dt},
\qquad
\Phi(q,t)=Ftq+\frac{F^2}{6m}t^3 .
\label{eq:L0-LLP-total-derivative}
\end{equation}
The same boundary term relates the classical on-shell actions and supplies the
Bargmann phase of the quantum transformation \cite{Robinett}.

The associated time-dependent canonical map is
\begin{equation}
Q=q+\frac{F}{2m}t^2,
\qquad
P=p+Ft,
\label{eq:canonical-QP-mass}
\end{equation}
and is generated by
\begin{equation}
G_2(q,P,t)=\left(q+\frac{F}{2m}t^2\right)P-Ftq-\frac{F^2}{6m}t^3 .
\label{eq:G2-mass}
\end{equation}
Indeed, $Q=\partial_PG_2$, $p=\partial_qG_2=P-Ft$, and
\begin{equation}
H_{\rm LP}(q,p)+\frac{\partial G_2}{\partial t}=\frac{P^2}{2m} .
\label{eq:transformed-H-free}
\end{equation}
Thus the constant-force system is regularly mapped to the free particle.

The same bridge admits an extended-phase-space formulation.  Promoting time to a canonical coordinate with conjugate
momentum \(p_t\), introduce
\begin{equation}
\mathcal C_{\rm LP}=p_t+\frac{p^2}{2m}+Fq\approx0,
\qquad
\mathcal C_0=P_t+\frac{P^2}{2m}\approx0 .
\label{eq:extended-FP-LP-constraints}
\end{equation}
The generating function gives $P_t=p_t-\partial_tG_2$, and
\begin{equation}
\mathcal C_0=\mathcal C_{\rm LP}.
\label{eq:extended-constraint-equivalence}
\end{equation}
The time-dependent affine map is therefore an ordinary canonical transformation
of the extended phase space, although its $t^2$ dependence makes the lifted map
polynomial rather than affine in all canonical variables.

At the quantum level, if $\Psi_0(Q,t)$ solves the free Schr\"odinger equation, then
\begin{equation}
\psi_{\rm LP}(q,t)=
\exp\left[-\frac{i}{\hbar}\left(Ftq+\frac{F^2}{6m}t^3\right)\right]
\Psi_0\left(q+\frac{F}{2m}t^2,t\right)
\label{eq:LP-TDSE-bridge-mass}
\end{equation}
solves $(i\hbar\partial_t-H_{\rm LP})\psi_{\rm LP}=0$.  This quantum affine bridge is a Weyl displacement with the Bargmann phase
\eqref{eq:L0-LLP-total-derivative}, rather than a quadratic metaplectic kernel.

\section{Propagator and singular oscillator limits}
\label{sec:LP-propagator-limits}

For $T=t_b-t_a>0$, the LP classical action is
\begin{equation}
S_{\rm LP}(q_b,t_b;q_a,t_a)
=
\frac{m(q_b-q_a)^2}{2T}
-\frac{FT}{2}(q_b+q_a)
-\frac{F^2T^3}{24m} .
\label{eq:LP-onshell-action}
\end{equation}
With $Q_j=q_j+Ft_j^2/(2m)$, $j=a,b$, and $\Phi$ from
\eqref{eq:L0-LLP-total-derivative}, it satisfies
\begin{equation}
S_{\rm LP}=S_0(Q_b,t_b;Q_a,t_a)-\Phi(q_b,t_b)+\Phi(q_a,t_a),
\qquad
S_0=\frac{m(Q_b-Q_a)^2}{2T} .
\label{eq:onshell-action-boundary-relation}
\end{equation}
Since the action is quadratic, the Van Vleck formula is exact:
\begin{equation}
K_{\rm LP}(q_b,t_b;q_a,t_a)
=
\left(\frac{m}{2\pi i\hbar T}\right)^{1/2}
\exp\left(\frac{i}{\hbar}S_{\rm LP}\right) .
\label{eq:LP-propagator}
\end{equation}
Equivalently, the affine bridge gives directly
\begin{equation}
K_{\rm LP}
=e^{-i\Phi(q_b,t_b)/\hbar}
K_0(Q_b,t_b;Q_a,t_a)
e^{i\Phi(q_a,t_a)/\hbar},
\qquad
K_0=\left(\frac{m}{2\pi i\hbar T}\right)^{1/2}
e^{im(Q_b-Q_a)^2/(2\hbar T)} .
\label{eq:kernel-affine-relation}
\end{equation}
Thus the Lagrangian boundary term, the wave-function intertwiner and the
propagator transformation are three forms of the same affine bridge.

\subsection{Displaced HO and IHO limits}
\label{subsec:oscillator-kernel-limits}

The oscillator routes are singular.  The two displaced Hamiltonians can be written
as
\begin{align}
H_{\omega,F}^{(+)}
&=
\frac{p^2}{2m}+\frac{m\omega^2}{2}
\left(q+\frac{F}{m\omega^2}\right)^2-\frac{F^2}{2m\omega^2}
=H_{\rm LP}+\frac{m\omega^2}{2}q^2,
\label{eq:displaced-HO}
\\
H_{\Omega,F}^{(-)}
&=
\frac{p^2}{2m}-\frac{m\Omega^2}{2}
\left(q-\frac{F}{m\Omega^2}\right)^2+\frac{F^2}{2m\Omega^2}
=H_{\rm LP}-\frac{m\Omega^2}{2}q^2.
\label{eq:displaced-IHO}
\end{align}
Their centers, $-F/(m\omega^2)$ and $F/(m\Omega^2)$, recede in opposite
directions as the corresponding frequency vanishes.

Let \(K_\nu^{(\pm)}\) denote the standard HO/IHO kernels, with
\(\nu=\omega\) for the upper sign and \(\nu=\Omega\) for the lower
sign, and with the usual Feynman--Maslov prescription understood in
the HO case:
\begin{equation}
K_\nu^{(\pm)}(y_b,y_a;T)
=
\left[\frac{m\nu}{2\pi i\hbar s_\pm(\nu T)}\right]^{1/2}
\exp\left\{
\frac{im\nu}{2\hbar s_\pm(\nu T)}
\left[(y_b^2+y_a^2)c_\pm(\nu T)-2y_by_a\right]
\right\},
\label{eq:HO-IHO-kernels-unified}
\end{equation}
where $(s_+,c_+)=(\sin,\cos)$ and $(s_-,c_-)=(\sinh,\cosh)$.  The displaced
kernels are
\begin{align}
K_{\omega,F}^{(+)}
&=
\exp\left(\frac{iF^2T}{2m\hbar\omega^2}\right)
K_\omega^{(+)}\left(q_b+\frac{F}{m\omega^2},
q_a+\frac{F}{m\omega^2};T\right),
\nonumber\\
K_{\Omega,F}^{(-)}
&=
\exp\left(-\frac{iF^2T}{2m\hbar\Omega^2}\right)
K_\Omega^{(-)}\left(q_b-\frac{F}{m\Omega^2},
q_a-\frac{F}{m\Omega^2};T\right).
\label{eq:shifted-oscillator-kernels}
\end{align}
The phases generated by the receding centers cancel the explicit constant
phases.  Denoting the corresponding classical actions by
\(S_{\omega,F}^{(+)}\) and \(S_{\Omega,F}^{(-)}\), one finds
\begin{equation}
S_{\omega,F}^{(+)}=S_{\rm LP}+O(\omega^2),
\qquad
S_{\Omega,F}^{(-)}=S_{\rm LP}+O(\Omega^2),
\qquad
\lim_{\omega\to0}K_{\omega,F}^{(+)}
=
\lim_{\Omega\to0}K_{\Omega,F}^{(-)}
=K_{\rm LP}.
\label{eq:oscillator-kernels-to-LP}
\end{equation}

The same limit acts on the dynamical integrals.  With
$y_+=q+F/(m\omega^2)$ and $y_-=q-F/(m\Omega^2)$,
\begin{align}
P_\omega^{(+)}&=p\cos\omega t+m\omega y_+\sin\omega t,
&
X_\omega^{(+)}&=y_+\cos\omega t-\frac{p}{m\omega}\sin\omega t,
\nonumber\\[-1mm]
P_\omega^{(+)}&\longrightarrow P_F,
&
X_\omega^{(+)}-\frac{F}{m\omega^2}&\longrightarrow X_F,
\nonumber\\[1mm]
P_\Omega^{(-)}&=p\cosh\Omega t-m\Omega y_-\sinh\Omega t,
&
X_\Omega^{(-)}&=y_-\cosh\Omega t-\frac{p}{m\Omega}\sinh\Omega t,
\nonumber\\[-1mm]
P_\Omega^{(-)}&\longrightarrow P_F,
&
X_\Omega^{(-)}+\frac{F}{m\Omega^2}&\longrightarrow X_F.
\label{eq:oscillator-integrals-limits}
\end{align}
Hence the Hamiltonians, dynamical integrals and propagators share one controlled
singular limit, distinct from the regular FP--LP accelerated-frame bridge.

\section{Airy transform and spectral condensation}
\label{sec:Airy-transform}

For definiteness take $F>0$; the case $F<0$ follows by reflection.  The stationary
Schr\"odinger equation
\begin{equation}
\left[-\frac{\hbar^2}{2m}\frac{d^2}{dq^2}+Fq\right]\psi_E=E\psi_E
\label{eq:LP-SSE}
\end{equation}
reduces to the Airy equation under
\begin{equation}
\kappa_F=\left(\frac{2mF}{\hbar^2}\right)^{1/3}=\ell_F^{-1},
\qquad
\xi=\kappa_F\left(q-\frac{E}{F}\right).
\label{eq:kappa-Airy-variable}
\end{equation}
The solution decaying for $q>E/F$ and normalized by
$\langle E|E'\rangle=\delta(E-E')$ is
\begin{equation}
\psi_E(q)=\frac{\kappa_F}{\sqrt F}
\operatorname{Ai}\left[\kappa_F\left(q-\frac{E}{F}\right)\right].
\label{eq:Airy-energy-eigenfunction}
\end{equation}
The spectrum is continuous and covers the real line.

\subsection{Cubic-phase transform from FP plane waves}
\label{subsec:cubic-Airy-transform}

In momentum representation, $q=i\hbar\,d/dp$, and the stationary equation becomes
\begin{equation}
\left(\frac{p^2}{2m}+i\hbar F\frac{d}{dp}\right)\widetilde\psi_E
=E\widetilde\psi_E,
\qquad
\widetilde\psi_E(p)=\frac{1}{\sqrt{2\pi\hbar F}}
\exp\left[\frac{i}{\hbar F}\left(\frac{p^3}{6m}-Ep\right)\right].
\label{eq:cubic-kernel}
\end{equation}
Fourier transformation gives
\begin{equation}
\psi_E(q)=\frac{1}{2\pi\hbar\sqrt F}
\int_{-\infty}^{+\infty}dp\,
\exp\left\{\frac{i}{\hbar}\left[
p\left(q-\frac{E}{F}\right)+\frac{p^3}{6mF}\right]\right\}.
\label{eq:Airy-plane-wave-superposition}
\end{equation}
With $p=\hbar\kappa_Fs$, this becomes the standard oscillatory representation
of $\operatorname{Ai}(\xi)$.  The transform is cubic-phase multiplication
followed by a Fourier transform in the conjugate variables \(p\) and \(E/F\),
normalized by \(\langle E|E'\rangle=\delta(E-E')\).

It should not be confused with the time-dependent affine intertwiner
\eqref{eq:LP-TDSE-bridge-mass}.  That map is a Weyl displacement relating the FP
and LP dynamics, whereas \eqref{eq:cubic-kernel} changes from the conserved free
momentum basis to the LP energy basis.  It is the spectral counterpart of the
affine extension and is not metaplectic: quadratic bridges carry quadratic
phases, while the LP spectral transform carries the cubic Airy phase.

\subsection{Hermite-to-Airy limit and condensation of the HO spectrum}
\label{subsec:Hermite-Airy-limit}

Let $\varphi_n^{(\omega)}$ be the normalized $n$-th eigenfunction of the harmonic
oscillator of frequency $\omega$.  The displaced-HO eigenfunctions are simply
\begin{equation}
\chi_{n,\omega}(q)=\varphi_n^{(\omega)}
\left(q+\frac{F}{m\omega^2}\right),
\qquad
E_n(\omega)=\hbar\omega\left(n+\frac12\right)-\frac{F^2}{2m\omega^2}.
\label{eq:shifted-Hermite-functions}
\end{equation}
A finite LP energy is obtained in the double scaling limit
\begin{equation}
\omega\to0,
\qquad
n\to\infty,
\qquad
E_n(\omega)\to E,
\qquad
n+\frac12
=\frac{F^2}{2m\hbar\omega^3}+\frac{E}{\hbar\omega}+o(\omega^{-1}).
\label{eq:HO-double-scaling}
\end{equation}
Since $E_{n+1}-E_n=\hbar\omega\to0$, the discrete HO levels condense into the
continuous LP spectrum.

The turning points are
\begin{equation}
q_\pm(E,\omega)=\frac{-F\pm\sqrt{F^2+2m\omega^2E}}{m\omega^2},
\qquad
q_+=\frac{E}{F}+O(\omega^2),
\qquad
q_-=-\frac{2F}{m\omega^2}-\frac{E}{F}+O(\omega^2).
\label{eq:HO-turning-points}
\end{equation}
Thus one turning point becomes the LP turning point while the other escapes to
$-\infty$. The fixed-\(F\) displaced-HO-to-Airy wave-function limit was analyzed
in Ref.~\cite{DaboulPogosyanWolf2007}.  Here it constitutes the
HO--LP spectral edge of the affine construction.  In the present
notation, the Plancherel--Rotach asymptotic formula
\cite{Olver,Szego1975} gives, after a consistent phase choice,
\begin{equation}
\frac{1}{\sqrt{\hbar\omega}}\,
\chi_{n(E,\omega),\omega}(q)
\longrightarrow
\frac{\kappa_F}{\sqrt F}
\operatorname{Ai}\left[\kappa_F\left(q-\frac{E}{F}\right)\right].
\label{eq:Hermite-to-Airy-limit}
\end{equation}
The factor $(\hbar\omega)^{-1/2}$ converts discrete normalization into
energy-delta normalization; the limit is understood on fixed compact $q$ intervals
or in the generalized spectral sense.

The same Airy eigenstate is therefore reached through three
complementary constructions: 
a regular cubic-phase change of spectral representation from FP plane
waves, a singular condensation limit of highly excited displaced-HO
states, and, as shown in
Appendix~\ref{app:IHO-LP-spectral-limit}, the limit of a subdominant
IHO parabolic-cylinder scattering branch.  The two oscillator routes
share the same local turning-point normal form but differ globally:
the HO spectrum condenses from discrete to continuous, whereas the
IHO spectrum is continuous before the limit and no level condensation
occurs.

\section{Crossed-field affine extension and Hall drift}
\label{sec:crossed-field-extension}

The LP system describes a charged particle in a homogeneous electric field
(\(F=-eE\) in one dimension).  Adding a perpendicular constant magnetic field
gives a planar extension: the pure Landau problem belongs to the quadratic
oscillator/CBT sector, while the electric field acts linearly on the magnetic
Heisenberg variables.

\subsection{Affine moving frames and the common origin of the interactions}
\label{subsec:affine-moving-frame-origin}

Consider the two-dimensional free particle in a time-dependent
orientation-preserving Euclidean frame.  With planar vectors understood as
column vectors, let
\begin{equation}
\mathbf R(t)=\mathbf b(t)+\mathcal R(\varphi(t))\mathbf r(t),
\quad
\mathcal R(\varphi)=e^{\varphi J},
\quad
J=
\begin{pmatrix}
0&-1\\
1&0
\end{pmatrix},
\quad
\Omega=\dot\varphi,
\quad
\mathbf u=\mathcal R^{-1}\dot{\mathbf b}.
\label{eq:affine-frame-data}
\end{equation}
Thus \(J\mathbf r=\hat{\mathbf z}\times\mathbf r\), with positive
\(\varphi\) corresponding to counter-clockwise rotation; below
\(\Omega\) is taken constant.  The inertial velocity resolved on the moving axes is the affine covariant velocity
\begin{equation}
\mathcal R^{-1}\dot{\mathbf R}
=
\nabla_t\mathbf r
:=
\dot{\mathbf r}+\Omega J\mathbf r+\mathbf u .
\label{eq:affine-covariant-velocity}
\end{equation}
Thus the free Lagrangian is $L_0=m|\nabla_t\mathbf r|^2/2$. After discarding the total derivative
\(d(m\mathbf u\cdot\mathbf r)/dt\) and the purely time-dependent term
\(m\mathbf u^2/2=m\dot{\mathbf b}^{\,2}/2\), which can likewise be
absorbed into a total derivative, one obtains 
\begin{equation}
L_{\rm mov}
\simeq
\frac{m}{2}\dot{\mathbf r}^{\,2}
+m\Omega(J\mathbf r)\cdot\dot{\mathbf r}
+\frac{m\Omega^2}{2}\mathbf r^2
-m\mathbf a\cdot\mathbf r,
\qquad
\mathbf a=\dot{\mathbf u}+\Omega J\mathbf u=\mathcal R^{-1}\ddot{\mathbf b} .
\label{eq:moving-frame-expanded-L}
\end{equation}
The four terms are the free kinetic, Coriolis/Landau, centrifugal and
homogeneous-force contributions.  For
\(\mathbf a=\mathbf a_0=\mathrm{const}\), the last is the LP, or homogeneous
electric-field, interaction.  The canonical momentum and Hamiltonian can be
written compactly as
\begin{align}
\mathbf p&=m\dot{\mathbf r}+m\Omega J\mathbf r,
\nonumber\\
H_{\rm mov}
&=
\frac{1}{2m}\bigl(\mathbf p-m\Omega J\mathbf r\bigr)^2
-\frac{m\Omega^2}{2}\mathbf r^2
+m\mathbf a\cdot\mathbf r
\nonumber\\
&=
\frac{\mathbf p^2}{2m}-\Omega L_z+m\mathbf a\cdot\mathbf r ,
\label{eq:moving-frame-Hamiltonian}
\end{align}
where \(L_z=\mathbf p\cdot J\mathbf r=xp_y-yp_x\).
Introduce the effective electromagnetic potentials and fields through
\begin{equation}
\frac{e}{c}\mathbf A_{\rm eff}=m\Omega J\mathbf r,
\qquad
\mathbf B_{\rm eff}=\frac{2mc}{e}\Omega\,\hat{\mathbf z},
\qquad
e\mathbf E=-m\mathbf a .
\label{eq:moving-frame-EM-identification}
\end{equation}
Identifying the magnetic field with the effective one,
\(\mathbf B\equiv\mathbf B_{\rm eff}\), define the corresponding
crossed-field Hamiltonian by
\[
H_{E,B}
=
\frac{1}{2m}
\left(
\mathbf p-\frac{e}{c}\mathbf A_{\rm eff}
\right)^2
-e\mathbf E\cdot\mathbf r .
\]
Then
\begin{equation}
H_{\rm mov}
=
H_{E,B}
-\frac{m\Omega^2}{2}\mathbf r^2,
\qquad
\omega_c\equiv\frac{eB}{mc}=2\Omega .
\label{eq:moving-frame-Landau-IHO}
\end{equation}

The Coriolis term linear in \(\Omega\) reproduces the orbital Larmor
coupling, while \eqref{eq:moving-frame-Landau-IHO} shows that the exact
rotating-frame Hamiltonian differs from the crossed-field Hamiltonian
\(H_{E,B}\) by the centrifugal \(O(\Omega^2)\) IHO potential.  Adding
the compensating isotropic HO term \(m\Omega^2\mathbf r^2/2\) therefore
yields \(H_{E,B}\); when \(\mathbf a=0\), equivalently
\(\mathbf E=0\), \(H_{E,B}\) reduces to the pure Landau Hamiltonian.
In this way FP, LP, IHO and Landau structures arise from the
translational, rotational and quadratic parts of one affine
moving-frame construction, with the HO supplying the compensation
required to isolate the Landau sector.

The rotational mechanism was analyzed in Ref.~\cite{InzunzaPlyushchay2022};
the accelerated translational sector in
\eqref{eq:moving-frame-expanded-L} adds the LP/electric term.  Here ``gauge''
means a local time-dependent Galilean frame: \eqref{eq:affine-covariant-velocity}
is a covariant, or elongated, derivative, not an electromagnetic $U(1)$ gauge
transformation.

\subsection{Crossed-field spectrum and guiding-center drift}
\label{subsec:guiding-center-affine}

Take $\mathbf E=E\hat{\mathbf x}$ and $\mathbf B=B\hat{\mathbf z}$, and choose the
Landau gauge $\mathbf A=(0,Bx,0)$, $\phi=-Ex$.  For a signed charge $e$,
\begin{equation}
H_{E,B}
=
\frac{p_x^2}{2m}
+
\frac{1}{2m}\left(p_y-\frac{eB}{c}x\right)^2
-eEx,
\qquad
\omega_c=\frac{eB}{mc}>0 .
\label{eq:crossed-Hamiltonian}
\end{equation}
Since $p_y=\hbar k$ is conserved, introduce
$x_k=\hbar k/(m\omega_c)$, $a_E=eE/(m\omega_c^2)$ and
$X_k=x_k+a_E$.  Completing the square gives
\begin{equation}
H_{E,B}(k)
=
\frac{p_x^2}{2m}
+
\frac{m\omega_c^2}{2}(x-X_k)^2
-eE x_k
-
\frac{(eE)^2}{2m\omega_c^2}.
\label{eq:crossed-completed-square}
\end{equation}
The corresponding generalized eigenfunctions and energy eigenvalues,
with \(n=0,1,\ldots\) and \(k\in\mathbb R\), are
\begin{align}
\Psi_{n,k}(x,y)
&=
\frac{e^{iky}}{\sqrt{2\pi}}\,
\varphi_n^{(\omega_c)}(x-X_k),
\nonumber\\
\mathcal E_{n,k}
&=
\hbar\omega_c\left(n+\frac12\right)
-
\frac{eE\hbar k}{m\omega_c}
-
\frac{(eE)^2}{2m\omega_c^2}.
\label{eq:crossed-spectrum}
\end{align}
Here \(\varphi_n^{(\omega_c)}\) is the normalized \(n\)-th eigenfunction
of the one-dimensional harmonic oscillator of frequency \(\omega_c\).
The corresponding group velocity
\(v_D=\hbar^{-1}\partial_k\mathcal E_{n,k}\) is independent of \(n\)
and \(k\), and
\begin{equation}
\mathbf v_D=v_D\hat{\mathbf y}
=
c\,\frac{\mathbf E\times\mathbf B}{B^2},
\qquad
v_D=-\frac{eE}{m\omega_c}=-\frac{cE}{B},
\label{eq:Hall-drift}
\end{equation}
which is the quantum counterpart of the classical Hall drift.  This
single-particle mechanism should not be confused with the many-body quantum Hall
effect.

The affine structure is most transparent in gauge-covariant variables.
In the symmetric gauge, define
\begin{equation}
\begin{gathered}
\pi_x=p_x+\frac{eB}{2c}y,
\qquad
\pi_y=p_y-\frac{eB}{2c}x,
\qquad
[\pi_x,\pi_y]=i\hbar m\omega_c,
\\
R_x=x+\frac{\pi_y}{m\omega_c},
\qquad
R_y=y-\frac{\pi_x}{m\omega_c},
\qquad
[R_i,\pi_j]=0,
\qquad
[R_x,R_y]=-\frac{i\hbar}{m\omega_c}.
\end{gathered}
\label{eq:guiding-center-algebra}
\end{equation}
The pure Landau Hamiltonian is $H_B=(\pi_x^2+\pi_y^2)/(2m)$, and the
noncommuting pair $(R_x,R_y)$ accounts for its degeneracy \cite{JohnsonLippmann1949}. Since
$x=R_x-\pi_y/(m\omega_c)$, the crossed-field Hamiltonian takes the form
\begin{equation}
H_{E,B}
=
\frac{\pi_x^2+(\pi_y-mv_D)^2}{2m}
-eE R_x
-
\frac{m v_D^2}{2}.
\label{eq:crossed-guiding-center-H}
\end{equation}
The electric field therefore shifts the cyclotron momentum and acts as a linear
potential on the noncommutative guiding-center plane.  The equations
$\dot R_x=0$ and $\dot R_y=v_D$ show that
$R_x$ and $R_y-v_Dt$ are dynamical integrals. This is the crossed-field counterpart of the LP pair:
a homogeneous force converts a conserved
center into a uniformly drifting one while preserving its Heisenberg structure.

\subsection{Affine intertwiner and extension of the Landau CBT}
\label{subsec:crossed-affine-intertwiner}
Returning to the Landau gauge of
\eqref{eq:crossed-Hamiltonian}, let
\(H_B=H_{E,B}|_{E=0}\).  Define
\(U_x(a)=\exp(-iap_x/\hbar)\), for which
\(U_x(a)xU_x(a)^{-1}=x-a\).  Then
\begin{equation}
H_{E,B}
=
U_x(a_E)H_B U_x(a_E)^{-1}
+v_Dp_y
-
\frac{m v_D^2}{2}.
\label{eq:crossed-affine-H-identity}
\end{equation}
Consequently, every solution $\Psi_B(t)$ of the pure Landau Schr\"odinger equation
generates a crossed-field solution
\begin{equation}
\Psi_{E,B}(t)
=
\exp\left(\frac{i m v_D^2t}{2\hbar}\right)
U_x(a_E)
\exp\left(-\frac{i}{\hbar}v_Dt\,p_y\right)
\Psi_B(t).
\label{eq:crossed-affine-wave-map}
\end{equation}
The three factors describe a scalar phase, a transverse displacement and a uniform
drift.  This is the planar crossed-field analogue of the affine FP--LP bridge.
If $S_B$ denotes the CBT from the appropriate two-dimensional free-particle data
to the pure Landau system \cite{InzunzaPlyushchayWipf2020}, then the
crossed-field bridge is simply
\begin{equation}
S_{E,B}(t)
=
\exp\left(\frac{i m v_D^2t}{2\hbar}\right)
U_x(a_E)
\exp\left(-\frac{i}{\hbar}v_Dt\,p_y\right)
S_B .
\label{eq:crossed-CBT-composition}
\end{equation}
Thus the pure-Landau CBT is supplemented by an affine displacement
and a uniform guiding-center drift.  Just as $H_{\rm LP}=H_0+Fq$ extends the quadratic
free-particle sector by a linear Heisenberg generator, the homogeneous electric
field extends the quadratic Landau/oscillator sector by a term linear in the
magnetic Heisenberg variables.  Fixing the cyclotron sector leaves precisely the
noncommutative guiding-center plane, on which the electric field acts as a linear
potential.

\section{Discussion and outlook}
\label{sec:discussion-outlook}

The LP system is an \emph{affine--Heisenberg companion} of the FP--HO--IHO
conformal triangle, not a fourth homogeneous quadratic vertex.  Its regular
FP--LP relation is a uniformly accelerated-frame map with a common boundary
function governing the canonical transformation, Bargmann phase and
propagator.  The HO--LP and IHO--LP relations instead require singular displaced
limits.  Spectrally, the Airy state arises from a cubic-phase transform of FP
momentum eigenstates, from condensation of highly excited HO levels, and from a
subdominant IHO scattering branch without level condensation.

The crossed-field extension provides a planar realization of the same
distinction.  Uniform acceleration generates the electric (LP) term,
uniform rotation the Coriolis/Landau coupling, and the squared
rotational connection the centrifugal IHO term; an HO
term compensates the latter to isolate the Landau sector.  The electric field
acts linearly on the noncommuting guiding-center plane, produces the Hall drift,
and supplements the Landau CBT by a displacement, drift and scalar phase.

Several extensions follow.  The guiding-center formulation points naturally to
the subcritical, supercritical and critical phases of the noncommutative Landau
problem, related by CBT and weak--strong duality
\cite{AlcalaPlyushchay2023}.  It remains to determine how the electric-field
affine intertwiner and Hall drift are realized across these phases.

A related relativistic direction starts from the identity of classical
evolution between a charge in a constant $(2+1)$-dimensional electromagnetic
field and the higher-derivative particle with torsion
\cite{Plyushchay1995TorsionEM}.  The ordinary nonrelativistic limit gives
crossed-field Landau dynamics, whereas the Jackiw--Nair limit leads to exotic
Galilean symmetry and noncommuting coordinates
\cite{DuvalHor,JackiwNair2000,HorvathyPlyushchay2002,HorvathyPlyushchay2004};
critical Hall dynamics and higher-derivative/Majorana formulations were studied
in Refs.~\cite{HorvathyPlyushchay2005,Plyushchay2006Majorana}.  Whether the
present planar system has an equivalent higher-derivative formulation requires
a separate off-shell and quantum analysis.

The same correspondence classifies the dual electromagnetic vector as timelike,
null or spacelike according as $B^2-\mathbf E^2$ is positive, zero or negative.
On the torsion side these are massive, massless and tachyonic Poincar\'e sectors
with elliptic, parabolic and hyperbolic classical evolution.  Geometric
quantization gives internal discrete-series representations $D_\alpha^\pm$ of
the universal cover of $SL(2,\mathbb R)$, while the physical constraint selects
a discrete massive mass--spin tower and generalized massless and tachyonic
states \cite{PlyuTorsion}.  Clarifying the relation of this relativistic
threefold structure to the Landau/HO, LP/Airy and IHO sectors remains open. 
The ordinary Galilean crossed-field problem arises from the
magnetic-dominated contraction, so no transition at
\(|\mathbf E|=|B|\) occurs within the nonrelativistic system considered
here.

Rindler kinematics provides a relativistic completion of the regular FP--LP
bridge: after subtracting the rest energy,
$H_{\rm R}=(1+aq/c^2)\sqrt{m^2c^4+c^2p^2}$ reduces to
$p^2/(2m)+maq+O(c^{-2})$ \cite{Rindler1966}.  Newton--Hooke mechanics arises
from dS/AdS contractions and null reductions of homogeneous pp waves
\cite{BacryLev,GibbonsPatricot2003}; more generally, the Bargmann--Eisenhart
lift embeds nonrelativistic mechanics into null geodesic dynamics
\cite{DuvalGibbonsHorvathy1991}.  Accelerated and rotating frames in $(A)dS$
may therefore provide a relativistic parent of $H_{\kappa,F}$, separating
curvature $\kappa$ from affine acceleration $F/m$; related horizon effects are
discussed in Ref.~\cite{DeserLevin1997}.

Another extension is suggested by de Alfaro--Fubini--Furlan conformal mechanics
\cite{deAlfaroFubiniFurlan1976} and the charge--monopole potential
$V(r)=m\omega^2r^2/2+\lambda/r^2$, with $\lambda=(eg)^2/(2m)$ in
Ref.~\cite{InzunzaPlyushchayWipf2020Monopole}. Since 
the \(r^{-2}\) interaction is homogeneous of degree \(-2\) and
preserves the underlying conformal-mechanics sector, 
one may ask whether the crossed-field construction admits
such a term, how it couples cyclotron and guiding-center sectors, and how it
modifies the Hall drift.  Finally, time-dependent translational and rotational
connections would lead to homogeneous nonstationary fields
\cite{LewRies}; their evolution operators should admit a factorization into
metaplectic, Heisenberg--Weyl and Bargmann-phase components within the Jacobi
group, with possible higher-dimensional, spin and curved-background extensions.
\section*{Acknowledgments}
The work was partially supported by the FONDECYT Project 1242046.

\appendix

\section{The IHO--LP spectral limit}
\label{app:IHO-LP-spectral-limit}

At the Hamiltonian, action and propagator levels, the IHO--LP limit follows from
the HO construction by the analytic continuation
\(\omega\mapsto\sigma i\Omega\), \(\sigma=\pm1\).  Spectrally, however,
normalizable Hermite states are replaced by generalized parabolic-cylinder
scattering states and an appropriate Stokes branch must be selected.

For the spectral problem, set \(y=q-F/(m\Omega^2)\) and
\(\varepsilon=E-F^2/(2m\Omega^2)\).  The shifted equation is
\[
\left[-\frac{\hbar^2}{2m}\frac{d^2}{dy^2}
      -\frac{m\Omega^2}{2}y^2\right]\chi_{E,\Omega}
=\varepsilon\chi_{E,\Omega}.
\]
Using \(D_n(z)=2^{-n/2}e^{-z^2/4}H_n(z/\sqrt2)\), its analytic continuation may be
written, up to normalization, as
\begin{equation}
\begin{aligned}
\chi_{E,\Omega}^{(\sigma)}(q)
&=
{\cal N}_{E,\Omega}^{(\sigma)}D_{\nu_\sigma}(z_\sigma),
&
 z_\sigma
&=
e^{\sigma i\pi/4}
\sqrt{\frac{2m\Omega}{\hbar}}
\left(q-\frac{F}{m\Omega^2}\right),
\\
\nu_\sigma+\frac12
&=
-\frac{\sigma i}{\hbar\Omega}
\left(E-\frac{F^2}{2m\Omega^2}\right).
\end{aligned}
\label{eq:app-IHO-parabolic-cylinder}
\end{equation}
The two signs select conjugate complex rotations and Stokes sectors.  This is an
analytic continuation of the differential equation, not a unitary continuation
of a normalized HO state: the integer label becomes a continuous complex index
and square-integrability is replaced by scattering conditions.

For \(F>0\), the potential
\(V_\Omega(q)=Fq-m\Omega^2q^2/2\) has turning points
\begin{equation}
q_-(E,\Omega)
=
\frac{E}{F}+O(\Omega^2),
\qquad
q_+(E,\Omega)
=
\frac{2F}{m\Omega^2}-\frac{E}{F}+O(\Omega^2),
\label{eq:app-IHO-turning-asymptotics}
\end{equation}
so that \(q_-\to E/F\), while \(q_+\to+\infty\).  Hence the finite turning point
becomes the LP turning point and the IHO forbidden interval tends to the LP
forbidden half-line \(q>E/F\).

The slope at the finite turning point is
\(s_\Omega=V_\Omega'(q_-)=\sqrt{F^2-2m\Omega^2E}\to F\).  With
\(\kappa_\Omega=(2ms_\Omega/\hbar^2)^{1/3}\) and
\(\xi_\Omega=\kappa_\Omega(q-q_-)\), the stationary equation takes the form
\begin{equation}
\frac{d^2\chi}{d\xi_\Omega^2}
=
\left(\xi_\Omega-\delta_\Omega\xi_\Omega^2\right)\chi,
\qquad
\delta_\Omega
=
\frac{m^2\Omega^2}{\hbar^2\kappa_\Omega^4}
\longrightarrow0.
\label{eq:app-IHO-local-equation}
\end{equation}
Consequently, on fixed turning-point intervals,
\(\xi_\Omega\to\kappa_F(q-E/F)\) and the parabolic-cylinder equation reduces to
the Airy equation.

A generic IHO scattering solution gives a local combination of
\(\operatorname{Ai}\) and \(\operatorname{Bi}\).  The LP eigenstate selected in
\eqref{eq:Airy-energy-eigenfunction} is obtained from the IHO branch that is
subdominant in the forbidden region.  Indeed, the barrier maximum,
\(q_{\rm max}=F/(m\Omega^2)\) and
\(V_{\rm max}=F^2/(2m\Omega^2)\), recedes to infinity and diverges as
\(\Omega\to0\); transmission is suppressed and the left-hand IHO scattering
problem becomes the totally reflected LP problem.  After removing a
\(q\)-independent scattering phase and matching the energy-delta normalization,
\begin{equation}
{\cal C}_{E,\Omega}\,
\chi_{E,\Omega}^{\rm sub}(q)
\longrightarrow
\frac{\kappa_F}{\sqrt F}
\operatorname{Ai}
\left[
\kappa_F\left(q-\frac{E}{F}\right)
\right].
\label{eq:app-IHO-to-Airy}
\end{equation}
The dominant barrier branch tends locally to \(\operatorname{Bi}\).  Unlike the
HO--LP limit, no condensation of discrete levels is involved: the IHO spectrum is
continuous for every \(\Omega>0\).  The two dashed spectral links therefore share
the same local Airy turning-point normal form but differ globally: the HO route is
a discrete-to-continuous large-level limit, whereas the IHO route is a
continuous-scattering limit with vanishing transmission.

\paragraph{Data Availability Statement.}
This article has no associated data or the data will not be deposited.

\paragraph{Code Availability Statement.}
This article has no associated code or the code will not be deposited.


\begin{thebibliography}{99}

\bibitem{Niederer1972}
U.~Niederer,
\emph{The maximal kinematical invariance group of the free Schr\"odinger equation},
\href{https://doi.org/10.5169/seals-114417}{Helv. Phys. Acta \textbf{45} (1972) 802--810}.

\bibitem{Hagen1972}
C.~R.~Hagen,
\emph{Scale and conformal transformations in Galilean-covariant field theory},
\href{https://doi.org/10.1103/PhysRevD.5.377}{Phys. Rev. D \textbf{5} (1972) 377--388}.

\bibitem{Niederer1973}
U.~Niederer,
\emph{The maximal kinematical invariance group of the harmonic oscillator},
\href{https://www.e-periodica.ch/cntmng?pid=hpa-001:1973:46::960}{Helv. Phys. Acta \textbf{46} (1973) 191--200},
\href{https://doi.org/10.5169/seals-114478}{doi:10.5169/seals-114478}.

\bibitem{InzunzaPlyushchayWipf2020}
L.~Inzunza, M.~S.~Plyushchay and A.~Wipf,
\emph{Conformal bridge between asymptotic freedom and confinement},
\href{https://doi.org/10.1103/PhysRevD.101.105019}{Phys. Rev. D \textbf{101} (2020) 105019},
\href{https://arxiv.org/abs/1912.11752}{arXiv:1912.11752}.

\bibitem{InzunzaPlyushchay2021Review}
L.~Inzunza and M.~S.~Plyushchay,
\emph{Conformal bridge transformation and $\mathcal{PT}$ symmetry},
\href{https://doi.org/10.1088/1742-6596/2038/1/012014}{J. Phys. Conf. Ser. \textbf{2038} (2021) 012014},
\href{https://arxiv.org/abs/2104.08351}{arXiv:2104.08351}.

\bibitem{AlcalaPlyushchay2026Projective}
A.~Alcala and M.~S.~Plyushchay,
\emph{Projective time, Cayley transformations and the Schwarzian geometry of the free particle--oscillator correspondence},
\href{https://doi.org/10.1103/ks12-b2v5}{Phys. Rev. D \textbf{113} (2026) 086007},
\href{https://arxiv.org/abs/2602.06378}{arXiv:2602.06378}.

\bibitem{AlcalaPlyushchay2026Triangle}
A.~Alcala and M.~S.~Plyushchay,
\emph{The free particle--oscillator--inverted oscillator triangle: conformal bridges, metaplectic rotations and $\mathfrak{osp}(1|2)$ structure},
\href{https://arxiv.org/abs/2605.09947}{arXiv:2605.09947}.

\bibitem{Bargmann}
V.~Bargmann,
\emph{On unitary ray representations of continuous groups},
\href{https://doi.org/10.2307/1969831}{Ann. Math. \textbf{59} (1954) 1--46}.

\bibitem{LeviLeb}
J.-M.~L\'evy-Leblond,
\emph{Galilei group and nonrelativistic quantum mechanics},
\href{https://doi.org/10.1063/1.1724319}{J. Math. Phys. \textbf{4} (1963) 776--788}.

\bibitem{BerryBalazs1979}
M.~V.~Berry and N.~L.~Balazs,
\emph{Nonspreading wave packets},
\href{https://doi.org/10.1119/1.11855}{Am. J. Phys. \textbf{47} (1979) 264--267}.

\bibitem{BernSch}
R.~Berndt and R.~Schmidt,
\emph{Elements of the Representation Theory of the Jacobi Group},
\href{https://doi.org/10.1007/978-3-0348-8772-4}
{Progress in Mathematics \textbf{163}, Birkh\"auser, 1998}.

\bibitem{DaboulWolf2006}
J.~Daboul and K.~B.~Wolf,
\emph{Noncommuting contractions of oscillators with constant force},
\href{https://doi.org/10.1088/0305-4470/39/16/005}
{J. Phys. A: Math. Gen. \textbf{39} (2006) 4173--4180}.

\bibitem{DaboulPogosyanWolf2007}
J.~Daboul, G.~S.~Pogosyan and K.~B.~Wolf,
\emph{Noncommuting limits of oscillator wave functions},
\href{https://doi.org/10.1134/S1063778807030106}
{Phys. At. Nucl. \textbf{70} (2007) 513--519}.

\bibitem{Robinett}
R.~W.~Robinett,
\emph{Quantum mechanical time-development operator for the uniformly accelerated particle},
\href{https://doi.org/10.1119/1.18179}{Am. J. Phys. \textbf{64} (1996) 803--808}.

\bibitem{Olver}
F.~W.~J.~Olver,
\emph{Uniform asymptotic expansions for Weber parabolic cylinder functions of large orders},
\href{https://doi.org/10.6028/jres.063B.014}{J. Res. Natl. Bur. Stand.  \textbf{63B} (1959) 131--169}.

\bibitem{Szego1975}
G.~Szeg\H{o},
\emph{Orthogonal Polynomials}, 4th ed.,  \href{https://bookstore.ams.org/coll-23}{American Mathematical Society
Colloquium Publications, Vol.~\textbf{23} (American Mathematical Society, Providence, 1975)}.

\bibitem{InzunzaPlyushchay2022}
L.~Inzunza and M.~S.~Plyushchay,
\emph{Dynamics, symmetries, anomaly and vortices in a rotating cosmic string background},
\href{https://doi.org/10.1007/JHEP01(2022)179}{JHEP \textbf{01} (2022) 179},
\href{https://arxiv.org/abs/2109.05161}{arXiv:2109.05161}.

\bibitem{JohnsonLippmann1949}
M.~H.~Johnson and B.~A.~Lippmann,
\emph{Motion in a constant magnetic field},
\href{https://doi.org/10.1103/PhysRev.76.828}
{Phys. Rev. \textbf{76} (1949) 828--832}.

\bibitem{AlcalaPlyushchay2023}
A.~Alcala and M.~S.~Plyushchay,
\emph{Weak--strong duality of the non-commutative Landau problem induced by a two-vortex permutation, and conformal bridge transformation},
\href{https://doi.org/10.1007/JHEP08(2023)141}{JHEP \textbf{08} (2023) 141},
\href{https://arxiv.org/abs/2304.06677}{arXiv:2304.06677}.

\bibitem{Plyushchay1995TorsionEM}
M.~S.~Plyushchay,
\emph{Relativistic particle with torsion and charged particle in a constant electromagnetic field: identity of evolution},
\href{https://doi.org/10.1142/S0217732395001587}{Mod. Phys. Lett. A \textbf{10} (1995) 1463--1469},
\href{https://arxiv.org/abs/hep-th/9309147}{arXiv:hep-th/9309147}.

\bibitem{DuvalHor}
C.~Duval and P.~A.~Horv\'athy,
\emph{The exotic Galilei group and the ``Peierls substitution''},
\href{https://doi.org/10.1016/S0370-2693(00)00341-5}{Phys. Lett. B \textbf{479} (2000) 284--290},
\href{https://arxiv.org/abs/hep-th/0002233}{arXiv:hep-th/0002233}.

\bibitem{JackiwNair2000}
R.~Jackiw and V.~P.~Nair,
\emph{Anyon spin and the exotic central extension of the planar Galilei group},
\href{https://doi.org/10.1016/S0370-2693(00)00379-8}{Phys. Lett. B \textbf{480} (2000) 237--238},
\href{https://arxiv.org/abs/hep-th/0003130}{arXiv:hep-th/0003130}.

\bibitem{HorvathyPlyushchay2002}
P.~A.~Horv\'athy and M.~S.~Plyushchay,
\emph{Non-relativistic anyons, exotic Galilean symmetry and noncommutative plane},
\href{https://doi.org/10.1088/1126-6708/2002/06/033}{JHEP \textbf{06} (2002) 033},
\href{https://arxiv.org/abs/hep-th/0201228}{arXiv:hep-th/0201228}.

\bibitem{HorvathyPlyushchay2004}
P.~A.~Horv\'athy and M.~S.~Plyushchay,
\emph{Anyon wave equations and the noncommutative plane},
\href{https://doi.org/10.1016/j.physletb.2004.05.043}{Phys. Lett. B \textbf{595} (2004) 547--555},
\href{https://arxiv.org/abs/hep-th/0404137}{arXiv:hep-th/0404137}.

\bibitem{HorvathyPlyushchay2005}
P.~A.~Horv\'athy and M.~S.~Plyushchay,
\emph{Nonrelativistic anyons in external electromagnetic field},
\href{https://doi.org/10.1016/j.nuclphysb.2005.02.027}{Nucl. Phys. B \textbf{714} (2005) 269--291},
\href{https://arxiv.org/abs/hep-th/0502040}{arXiv:hep-th/0502040}.

\bibitem{Plyushchay2006Majorana}
M.~S.~Plyushchay,
\emph{Majorana equation and exotics: higher derivative models, anyons and noncommutative geometry},
Electron. J. Theor. Phys. \textbf{3}, No.~10 (2006) 17--31,
\href{https://arxiv.org/abs/math-ph/0604022}{arXiv:math-ph/0604022}.

\bibitem{PlyuTorsion}
M.~S.~Plyushchay,
\emph{The model of the relativistic particle with torsion},
\href{https://doi.org/10.1016/0550-3213(91)90555-C}{Nucl. Phys. B \textbf{362}  (1991)  54--72}.

\bibitem{Rindler1966}
W.~Rindler,
\emph{Kruskal space and the uniformly accelerated frame},
\href{https://doi.org/10.1119/1.1972547}{Am. J. Phys. \textbf{34} (1966) 1174--1178}.

\bibitem{BacryLev}
H.~Bacry and J.-M.~L\'evy-Leblond,
\emph{Possible kinematics},
\href{https://doi.org/10.1063/1.1664490}{J. Math. Phys. \textbf{9} (1968) 1605--1614}.

\bibitem{GibbonsPatricot2003}
G.~W.~Gibbons and C.~E.~Patricot,
\emph{Newton--Hooke spacetimes, Hpp-waves and the cosmological constant},
\href{https://doi.org/10.1088/0264-9381/20/23/016}{Class. Quantum Grav. \textbf{20} (2003) 5225--5239},
\href{https://arxiv.org/abs/hep-th/0308200}{arXiv:hep-th/0308200}.

\bibitem{DuvalGibbonsHorvathy1991}
C.~Duval, G.~W.~Gibbons and P.~A.~Horv\'athy,
\emph{Celestial mechanics, conformal structures, and gravitational waves},
\href{https://doi.org/10.1103/PhysRevD.43.3907}
{Phys. Rev. D \textbf{43} (1991) 3907--3922},
\href{https://arxiv.org/abs/hep-th/0512188}{arXiv:hep-th/0512188}.

\bibitem{DeserLevin1997}
S.~Deser and O.~Levin,
\emph{Accelerated detectors and temperature in (anti-)de Sitter spaces},
\href{https://doi.org/10.1088/0264-9381/14/9/003}{Class. Quantum Grav. \textbf{14} (1997) L163--L168},
\href{https://arxiv.org/abs/gr-qc/9706018}{arXiv:gr-qc/9706018}.

\bibitem{deAlfaroFubiniFurlan1976}
V.~de~Alfaro, S.~Fubini and G.~Furlan,
\emph{Conformal invariance in quantum mechanics},
\href{https://doi.org/10.1007/BF02785666}{Nuovo Cim. A \textbf{34} (1976) 569--612}.

\bibitem{InzunzaPlyushchayWipf2020Monopole}
L.~Inzunza, M.~S.~Plyushchay and A.~Wipf,
\emph{Hidden symmetry and (super)conformal mechanics in a monopole background},
\href{https://doi.org/10.1007/JHEP04(2020)028}{JHEP \textbf{04} (2020) 028},
\href{https://arxiv.org/abs/2002.04341}{arXiv:2002.04341}.

\bibitem{LewRies}
H.~R.~Lewis and W.~B.~Riesenfeld,
\emph{An exact quantum theory of the time-dependent harmonic oscillator
and of a charged particle in a time-dependent electromagnetic field},
\href{https://doi.org/10.1063/1.1664991}{J. Math. Phys. \textbf{10} (1969) 1458--1473}.

\end{thebibliography}
\end{document}